\documentclass[pdftex,twocolumn,epjc3]{svjour3}          % twocolumn

\RequirePackage[T1]{fontenc}

\smartqed  % flush right qed marks, e.g. at end of proof

\RequirePackage{graphicx}
\RequirePackage{mathptmx}      % use Times fonts if available on your TeX system
\RequirePackage{flushend}
\RequirePackage[numbers,sort&compress]{natbib}
\RequirePackage[colorlinks,citecolor=blue,urlcolor=blue,linkcolor=blue]{hyperref}
\RequirePackage{amsmath}      % for multi-line equations

\begin{document}

\title{A fit-less approach to the elasticity of the handles in optical tweezers experiments} %\thanksref{t1}}


\author{Alessandro Mossa\thanksref{e1,addr1,addr2}
        \and
        Ciro Cecconi\thanksref{e2,addr3,addr4} %etc.
}

%\thankstext[$\star$]{t1}{Thanks to the title}
\thankstext{e1}{e-mail: alessandro.mossa@fi.infn.it}
\thankstext{e2}{e-mail: ciro.cecconi@gmail.com}

\institute{INFN Firenze, via Sansone 1, 50019 Sesto Fiorentino, Italy\label{addr1}
	\and
	ISIS ``Leonardo da Vinci'', via del Terzolle 91, 50127 Firenze, Italy\label{addr2}
          \and
          Department of Physics, Informatics and Mathematics, University of Modena and Reggio Emilia, via Giuseppe Campi 213/a, 41125 Modena, Italy\label{addr3}
          \and
         Center S3, CNR Institute Nanoscience, via Giuseppe Campi 213/a, 41125 Modena, Italy\label{addr4}
}

\date{Received: date / Accepted: date}

\abstractdc{The elastic properties of the double-stranded DNA handles used in optical tweezers experiments on biomolecules are customarily modeled by an extensible worm-like chain model. Fitting such model to experimental data however is no trivial task, as the function depends on four parameters in a highly non-linear fashion. We hereby propose a method to bypass the fitting procedure and obtain an empirical force \textit{vs.}\ extension curve that accurately reproduce the elasticity of the handles.}

\maketitle

\section{Introduction}

\begin{figure}
\begin{minipage}{\columnwidth}
\centering
%\framebox[\columnwidth][c]{\raisebox{0pt}[20mm][20mm]{figure image}}
\includegraphics[width=\textwidth]{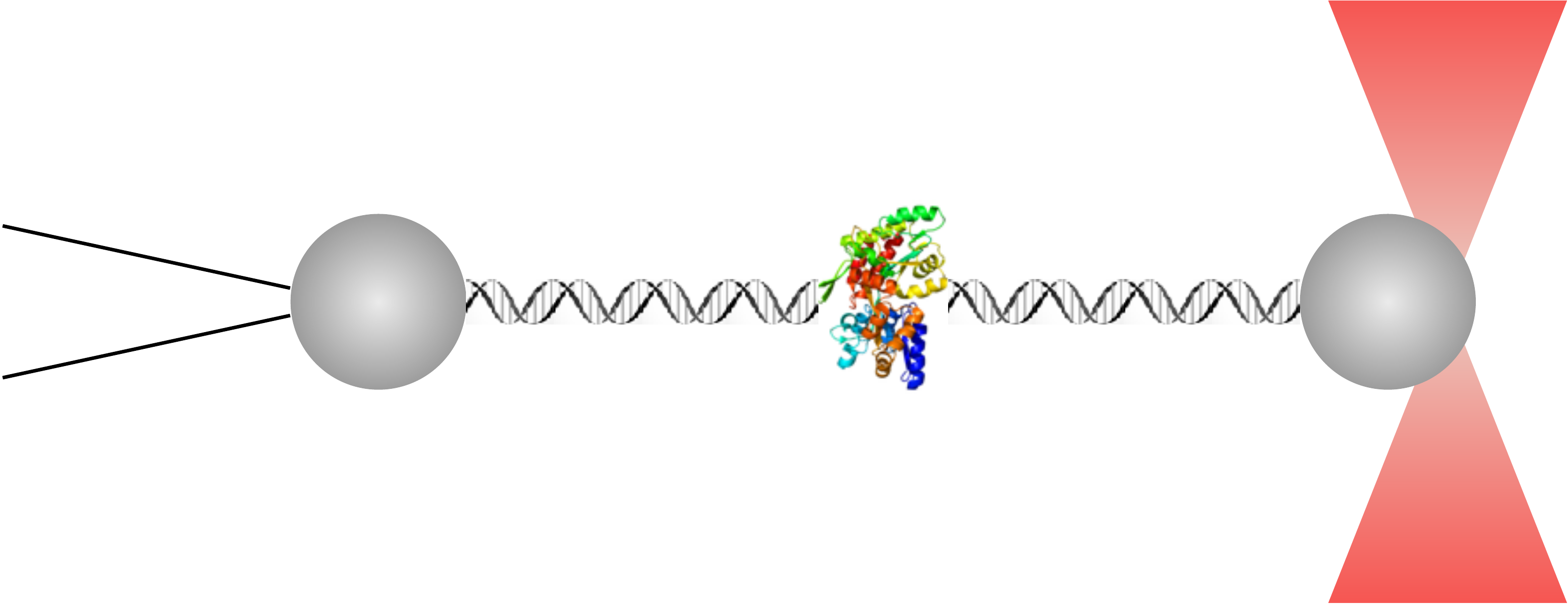}
\end{minipage}
\caption{Experimental setting: one bead (to the left) is held fixed to the end of a pipette, 
the other one (to the right) is optically trapped. A protein is tethered to the beads by means 
of two double-stranded DNA handles. \label{fig1}}
\end{figure}

Optical tweezers are nowadays one of the main experimental techniques used in single-molecule biophysics \cite{ref:OT}. Proteins or nucleic acids can be manipulated by force in the piconewton scale by exploiting the ability of focused laser light to trap micrometer-sized polystirene beads. In order to avoid non-specific interactions, the molecule of interest is tethered to the beads by means of a pair of double-stranded DNA handles (see fig.~\ref{fig1}), whose length is usually chosen in the range 20--800 base pairs. 

Among the various equilibrium and non-equilibrium experiments that can be performed within this setup, here we are concerned with the force-ramp protocol, where the distance between the trap and a reference point (which is usually either the tip of a pipette, or another optical trap) is increased (stretching phase) or decreased (relaxation phase) at a constant speed. Throughout the process the force that the trap exerts on the trapped bead and the distance between the beads are measured. The force \textit{vs.}\ extension curve (shortened to FEC henceforth) obtained in this way is dominated by the elastic behaviour of the handles, which one needs to estimate and subtract during data analysis in order to get useful information about the molecule in between. 

The most common approach is to model the elasticity of the handles with an extensible worm-like chain \cite{ref:eWLC}. This requires a non-linear fit with four parameters, which can (actually, very often does) give poor agreement with the experimental data, or assign very unlikely values to the parameters. In the following, we will illustrate the shortcomings of this standard recipe, and propose a novel method to deal with handles elasticity which does not depend on any fitting procedure. The first step is to build an empirical FEC by combining data belonging to the low-force regime of a stretching trace with data from the high-force regime of a relaxation trace. From such baseline a reference grid of FECs can be constructed and employed to ease the data analysis process.    

\begin{figure} 
\begin{minipage}{\columnwidth}
\centering
%\framebox[\columnwidth][c]{\raisebox{0pt}[20mm][20mm]{figure image}}
\includegraphics[width=\textwidth]{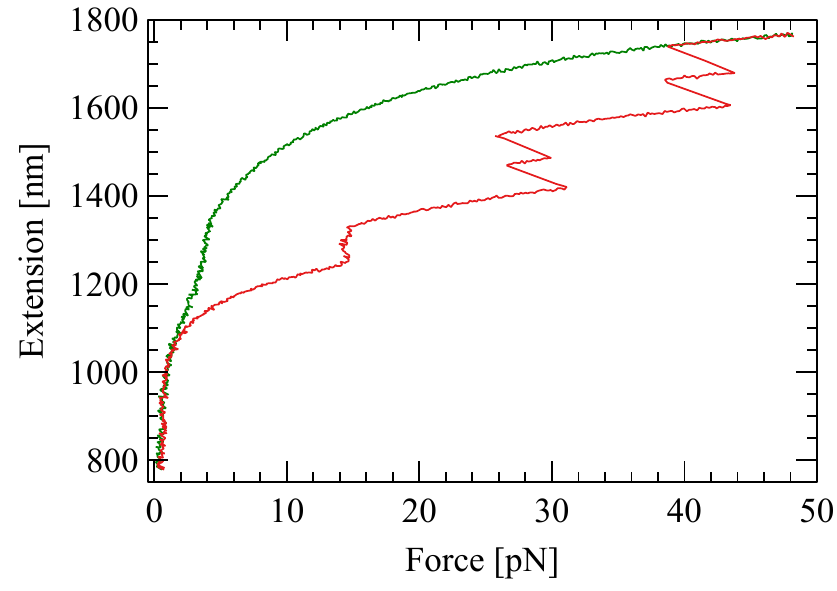}
\end{minipage}
\caption{Stretching (red trace) and subsequent relaxation (green trace) of a homotetramer form of MBP.  \label{fig2}}
\end{figure}

To make the presentation more definite, we will refer to a specific example, taken from a recent experiment \cite{ref:HSPB8}. In the concrete application we describe, the system under scrutiny is a construct comprised of four repeats of maltose binding protein (MBP for short), a 370-residue protein \cite{ref:MBPstr} characterized by a 279-residue core and 5 peripheral alpha helices \cite{ref:MBPtans}. When the stretching starts from the native state of the homotetramer\footnote{That may happen only in the first pulling with any given molecule: subsequent cycles are affected by aggregation phenomena, which is the very reason why this construct has been engineered to begin with.}, the FEC of a stretching-relaxation cycle typically resembles fig.~\ref{fig2}. We clearly distinguish four \emph{jumps}: sudden drops in the force accompanied by an increase of the extension. They signal the denaturation of the four core structures. The more gradual increase in the extension at a force around 14 pN is due to the unfolding of the alpha helices. 

The relaxation trace, in comparison, looks less eventful. A gradual compaction of the peptide chain starts to take place around 5 pN; before that, the curve follows the elastic behaviour of the whole molecular construct comprising the handles and the completely denatured MBP tetramer. 

An important point to note about the FECs is that the extension is actually measured with respect to an unknown additive constant. This adds one more parameter to the fitting procedure we need to carry out when we try to describe experimental data by means of the polymer elasticity models discussed in the next section.

\section{Elastic models}

When considering the response to tension of the molecular construct depicted in fig.~\ref{fig1}, it is useful to divide the distance $x$ between the beads into three contributions:
\begin{equation}
	x = x_\mathrm{h}+x_\mathrm{f}+x_\mathrm{u} \,,
\end{equation}  
where $x_\mathrm{h}$, $x_\mathrm{f}$, $x_\mathrm{u}$ are the extensions of the handles, of the folded portion of the protein, and of the unfolded peptide chain, respectively.

The equilibrium elastic behaviour of the folded protein $x_\mathrm{f}(F)$ at force $F$ may be modeled by a freely-jointed chain of a polymer with identical Kuhn and contour lengths \cite{ref:xf}:
\begin{equation} \label{eq:FJC}
	x_\mathrm{f}(F) = d\left[ \coth\left(\frac {F d}{k_\mathrm{B}T}\right)-\frac{k_\mathrm{B}T}{F d} \right] \,,
\end{equation}
where $k_\mathrm{B}$ is Boltzmann's constant, $T$ is the absolute temperature and $d$ is the contour length of the folded protein, which can be estimated from the structures deposited in the Protein Data Bank \cite{ref:PDB} as $d=(3.99\pm 0.01)$ nm (reduced to $d=(1.756\pm0.003)$ nm for the core structure alone). 

Following a well established practice, the equilibrium elastic properties of the unfolded peptide chain are modeled by a Marko--Siggia \cite{ref:MS} worm-like chain (WLC for short) with the correction terms computed by Bouchiat \textit{et al.}\ \cite{ref:MSB}
\begin{equation} \label{eq:WLC}
	F(x_\mathrm{u}) = \frac{k_\mathrm{B}T}{L_\mathrm{p}}\left[ \frac{1}{4(1-\ell)^2}-\frac{1}{4}+\ell+\sum_{i=2}^7 a_i \ell^i \right]\,,
\end{equation}
with $\ell = x/L_\mathrm{c}$ and
\begin{align*}
	a_2 &= -0.5164228 &\quad a_3 &= -2.737418 \\ 
	a_4 &= 16.07497 &\quad  a_5 &=  -38.87607 \\ 
	a_6 &= 39.49944 &\quad a_7 &= -14.17718 \,.
\end{align*}
The choice of parameters is the same that has been successfully adopted in other past experiments \cite{ref:Barn,ref:ACBP}, that is persistence length $L_\mathrm{p}=0.65$ nm and contour length $L_\mathrm{c}=nl_0$, where the inter-residue distance $l_0=0.34$ nm is multiplied by the number $n$ of unfolded amino acids\footnote{Actually, in the cited references we used $L_\mathrm{p} = 0.60$ nm: the larger value was chosen here to optimize fig.~\ref{fig4}, as explained towards the end of sec.~\ref{sec3}.}. 

The two models in eqs.~(\ref{eq:FJC}) and (\ref{eq:WLC}) only consider entropic effects, which is adequate for typical optical tweezers experiments. The DNA handles, however, get usually stretched to the point that it is no longer possible to disregard enthalpic effects. It is customary to deal with them by modifying the WLC into an extensible version of it (eWLC in the following) \cite{ref:eWLC} which is still written as eq.~(\ref{eq:WLC}) with the understanding that
\begin{equation}
	\ell = \frac{x}{L_\mathrm{c}}-\frac{F}{Y} \,,
\end{equation}
where $Y$ is the Young modulus. Modeling the handles elasticity with an eWLC therefore requires a highly non-linear fitting procedure over four parameters: $L_\mathrm{p}$, $L_\mathrm{c}$, $Y$ and $x_0$ because, as we have already remarked, the measured extension is shifted with respect to the physical extension by an unknown additive term. 

It is easy to realize what can go wrong with such fit. With reference to fig.~\ref{fig2}, the range of data where only the elastic effect of the handles is relevant is restricted to the region below 14 pN, which is precisely the region where the effect of the enthalpic correction is negligible, so we can expect to have huge uncertainty in the determination of $Y$. However, the Young modulus so poorly determined is essential when we extrapolate the theoretical curve for handles elasticity to the high-force region, where it is actually more useful. Fortunately, as we will show in the next section, there is a way to bypass the fit altogether.

\section{Building a baseline, then a grid} \label{sec3}       

While fitting the eWLC to handles elasticity is tricky, the model in eq.~\ref{eq:WLC} does a very good job of describing the elastic behaviour of the unfolded peptide chain. So good that we can actually use it to subtract its contribution from the relaxation trace in fig.~\ref{fig2}, thus obtaining an empirical curve that depends only on the behaviour of the handles, at least in the region before compaction begins. 

\begin{figure} 
\begin{minipage}{\columnwidth}
\centering
%\framebox[\columnwidth][c]{\raisebox{0pt}[20mm][20mm]{figure image}}
\includegraphics[width=\textwidth]{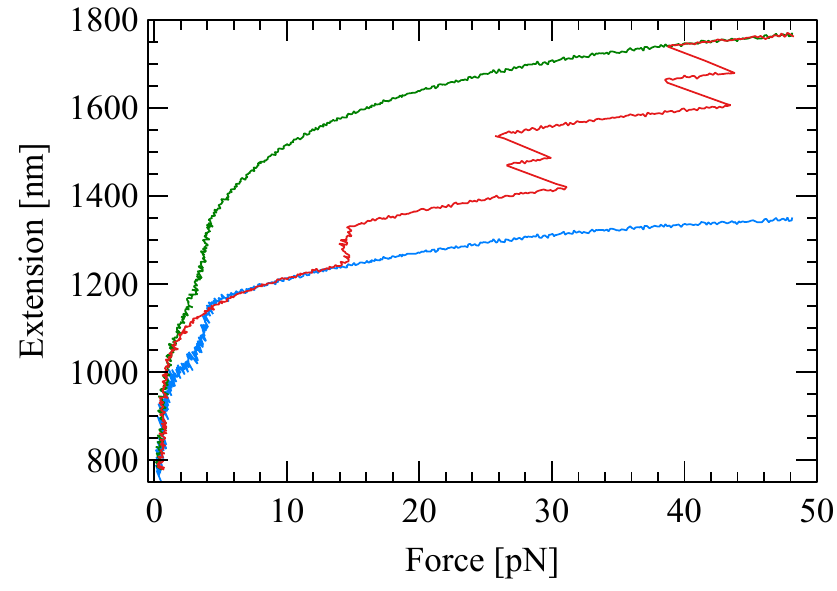}
\end{minipage}
\caption{Stretching (red trace) and subsequent relaxation (green trace) of a homotetramer form of MBP. It is also shown (in blue) the shifted relaxation curve defined by eq.~(\ref{eq:shift}). \label{fig3}}
\end{figure}

Let $x_\mathrm{r}(F)$ denote the experimental relaxation trace. We know that at very high forces the tetramer is completely denatured, so it behaves like a WLC with $L_\mathrm{p}=0.65$ nm and $L_\mathrm{c}=512.38$ nm (the total number of residues is 1507, including 27 amino acids employed as linkers between MBP repeats). Let $x_\mathrm{u}(F)$ be the function defined by numerical inversion of eq.~(\ref{eq:WLC}) (see \ref{app}); we can compute a shifted relaxation curve
\begin{equation} \label{eq:shift}
	x_\mathrm{s}(F) \equiv x_\mathrm{r}(F) - x_\mathrm{u}(F) 
\end{equation}      
and compare it to the stretching trace, as done in fig.~\ref{fig3}. The reasons why we are disregarding in this analysis the elastic properties of the folded MBP repeats given by eq.~(\ref{eq:FJC}) are discussed towards the end of this section.  

The striking feature of fig.~\ref{fig3} is the overlapping in the range of forces from 6 pN to 14 pN between the shifted relaxation curve (in colour blue) of eq.~(\ref{eq:shift}) and the stretching trace (in colour red). This suggests a simple recipe to deal with handles elasticity: if we join the low-force portion of the stretching trace, which corresponds to a phase where the protein is fully folded and no structure has been denatured yet, with the high-force portion of the shifted relaxation trace, which corresponds to a phase where the protein is fully unfolded and no structure as been formed yet, we obtain what we call the \emph{baseline}, an empirical reference curve that describes the elastic properties of the handles alone.           

Once the baseline is built, we can smooth it by applying a suitable filter, or we can fit an eWLC model to it: the already mentioned drawbacks of the fitting procedure are mitigated by the fact that now we are working with data that reach into the high-force regime. Either way, using the function $x_\mathrm{u}(f)$ we can add back to the baseline the elastic behaviour of an $n$-residue peptide chain, with $n$ going from 0 to the total size of the protein (1507 in our specific example), thus obtaining a reference grid of curves that can be used to reveal the presence of intermediate structures, and estimate their size. 

\begin{figure} 
\begin{minipage}{\columnwidth}
\centering
%\framebox[\columnwidth][c]{\raisebox{0pt}[20mm][20mm]{figure image}}
\includegraphics[width=\textwidth]{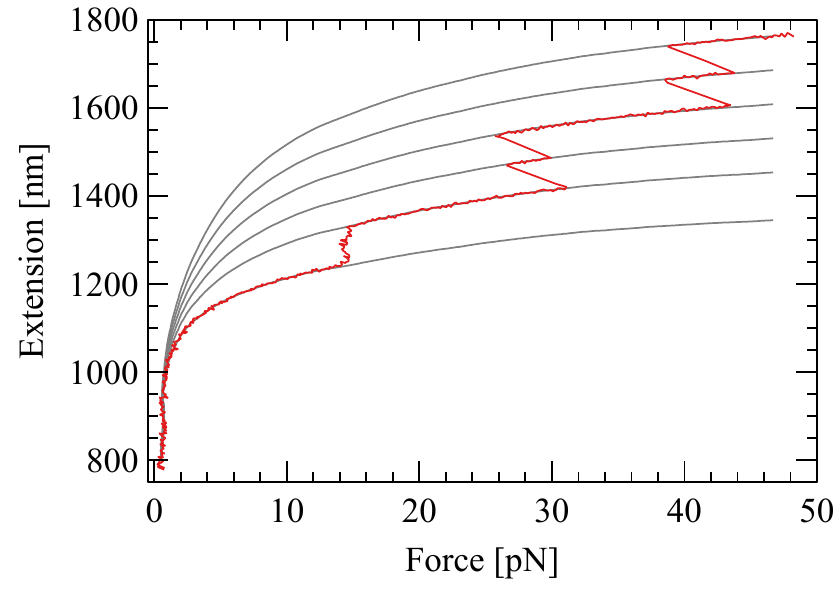}
\end{minipage}
\caption{In colour red, the stretching trace of a homotetramer form of MBP. In gray are shown the grid reference lines that correspond to  (from top to bottom) completely denatured molecule, 1, 2, 3, 4 surviving cores, or the intact native state. \label{fig4}}
\end{figure}

The use of a tetramer construct allows to perform a neat consistency check of the whole approach: fig.~\ref{fig4} shows the stretching trace of fig.~\ref{fig2} plotted against the grid lines corresponding to (from top to bottom) the wholly denatured protein ($n=1507$), one to four core structures still folded ($n=1228$, 949, 670, 391), and the completely folded native structure ($n=0$). The excellent agreement between the computed lines and the experimental trace must be interpreted as a statement about the goodness of the choice made for the parameters in eq.~(\ref{eq:WLC}). 

\begin{figure} 
\begin{minipage}{\columnwidth}
\centering
%\framebox[\columnwidth][c]{\raisebox{0pt}[20mm][20mm]{figure image}}
\includegraphics[width=\textwidth]{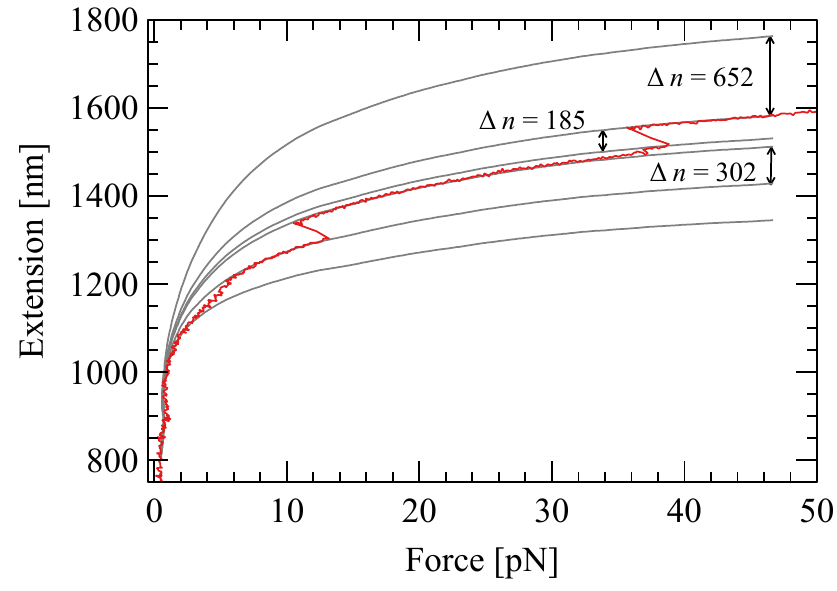}
\end{minipage}
\caption{In colour red, the stretching trace of a homotetramer form of MBP. In gray are shown the grid reference lines that have been used to estimate the size of the structures identified in the text. \label{fig5}}
\end{figure}

The same grid can be used to analyze all the experimental traces obtained from the same individual molecule; see for instance fig.~\ref{fig5} where the grid built for this example so far allows to recognize and size up in another stretching trace belonging to the same dataset the presence of a core that breaks around 14 pN, a smaller structure that breaks at about 40 pN, and an aggregate that remains unbroken at very high forces.   

One could argue that instead of eq.~(\ref{eq:shift}) the proper subtraction should be 
\begin{equation}
	x_\mathrm{s}(f) \equiv x_\mathrm{r}(f) - x_\mathrm{u}(f) + x_\mathrm{f}(f)\,,
\end{equation} 
taking into account also the contribution of the folded structures when they are present. After all, four MBP proteins in the native state at 12 pN tension are barely shorter than 16 nm, a shift that would be clearly visible, \textit{e.g.}\ in fig.~\ref{fig4}. The point though is that, while in a cycle that starts from the native state it is easy to know with a certain confidence what is folded and what is not at various stages of the experiment, this is impossible for a generic trace like the one in fig.~\ref{fig5}. In a pragmatic perspective, it is more useful to adjust the parameters for the WLC model of the peptide chain ignoring the folded structures' constribution: the small error is absorbed into the choice of the elastic model for the unfolded protein and we can use the grid to study any trace.    
  
\section{Conclusion}

In accord with the principle of making science as open and reproducible as possible, the data and the program used to produce the figures for this paper are publicly available in a GitHub repository \cite{ref:gh}. The interested readers can easily adapt the Python code to apply the baseline and grid method to their own data.

\begin{figure} 
\begin{minipage}{\columnwidth}
\centering
%\framebox[\columnwidth][c]{\raisebox{0pt}[20mm][20mm]{figure image}}
\includegraphics[width=\textwidth]{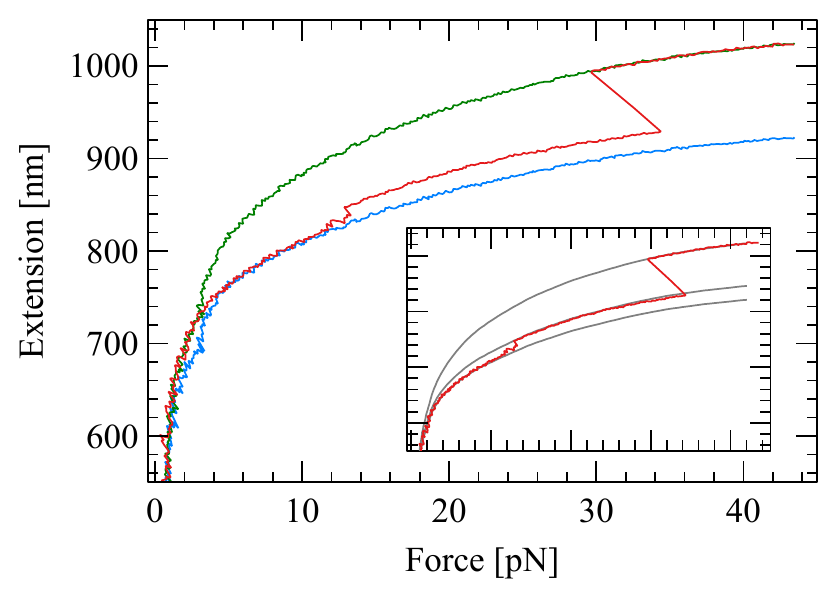}
\end{minipage}
\caption{Stretching (red trace) and subsequent relaxation (green trace) of a monomer form of MBP. It is also shown (in blue) the shifted relaxation curve defined by eq.~(\ref{eq:shift}). In the inset, the gray lines are the reference curves that correspond to (from top to bottom) completely denatured molecule, one core structure, or the intact native state. \label{fig6}}
\end{figure}

We have seen that the elastic behaviour of the handles can be represented by an empiric baseline obtained from experimental data without the need to fit the eWLC model. Also, a grid of reference curves can be built on top of the baseline and used to identify and size up intermediate states along the stretching or relaxation curves. It remains to be discussed how general is this method, and what are its limitations. First of all, there is nothing strictly peculiar to the tetramer example that has been used so far; fig.~\ref{fig6} demonstrates the effectiveness of the baseline and grid construction in the case of data obtained with a single MBP, while the Jupyter Notebook available as an online resource for this paper \cite{ref:gh} contains examples about other molecules as well. 

Of course, the more details we know about the protein that is being pulled, the more possibilities we have of setting nontrivial consistency checks like the one in fig.~\ref{fig4}. While the general idea is still valid, it's more difficult to be completely confident in the baseline built for a molecule about which one knows nothing except the structure. Another limitation that comes to mind is that for some systems it might be impossible to achieve the overlap between the stretching trace and the shifted relaxation trace shown in figs.~\ref{fig3} and \ref{fig6}: this is the case if unfolding in the stretching curve starts at a force for which refolding is already under way in the relaxation curve.

Apart from such peculiar cases, however, we think that the approach we have introduced is a definite improvement over the standard practices in the field of optical tweezers force-ramp experiments. It can be used either to provide a better sample for an eWLC fit, or to avoid the fit entirely. In any way, it provides a reliable method to factor out the elasticity of the handles, allowing to focus the attention on the properties of the system we are actually interested in.     

\begin{acknowledgements}
C.C.\ gratefully acknowledges the University of Modena and Reggio Emilia for financial support through the grant 020145\_17\_FDA\_CARRAFAR2016INTER.  
\end{acknowledgements}

\appendix

\section{An inversion formula for the WLC}\label{app}

The original Marko--Siggia formula for the WLC model  
\begin{equation} \label{eq:WLC0}
	f(\ell) = \frac{1}{4(1-\ell)^2}-\frac{1}{4}+\ell \,,
\end{equation}
with $\ell = x/L_\mathrm{c}$ and $f=F L_\mathrm{p}/(k_\mathrm{B}T)$ can actually be inverted explicitly by solving a cubic equation. The solution is given by Cardano's formula 
\begin{multline} \label{eq:Cardano}
	\ell = \frac{4f+9}{12}+\sqrt[3]{-\frac{\delta(f)+108}{1728}+\frac{\sqrt{\delta(f)}}{48\sqrt{3}}}+\\
	+\sqrt[3]{-\frac{\delta(f)+108}{1728}-\frac{\sqrt{\delta(f)}}{48\sqrt{3}}}
\end{multline}
if $f\leq f_0$ and by Vi\`ete's formula 
\begin{equation} \label{eq:Viete}
	\ell = \frac{4f+9}{12}-\frac{4f-3}{6}\cos\left[ \frac{\pi}{6}-\frac{1}{3}\arcsin \frac{\delta(f)+108}{(4f-3)^3} \right]
\end{equation}
if $f>f_0$, with $f_0=3(1+\sqrt[3]{4})/4$ and
\begin{equation}
	\delta(f) = -64f^3+144 f^2-108 f+135 \,.
\end{equation}
This may be used to provide a very good initial estimate to a numerical routine for inverting eq.~(\ref{eq:WLC}), as is done in the online resource that complement this paper \cite{ref:gh}, or to help setting up a fit routine. In this regard, note that the extensible version of eq.~(\ref{eq:WLC0}), which is obtained by replacing $\ell$ with $\tilde{\ell}$ defined as
\begin{equation}
	\tilde{\ell} = \ell-k f \,, \qquad k\equiv \frac{k_\mathrm{B}T}{L_\mathrm{p}Y}\,,
\end{equation}   
can also be inverted in the same way, yielding 
\begin{equation}
	\tilde{\ell} = \ell + k f
\end{equation} 
where $\ell$ is given by eqs.~(\ref{eq:Cardano}) and (\ref{eq:Viete}).

\section*{Author contribution statement}

A.M. devised the method here described while analyzing data from C.C.'s experiments. Both authors wrote the paper.

\end{document}